\documentclass[aps,prl,twocolumn,groupedaddress,floats,showpacs]{revtex4}

\usepackage{amsmath}
\usepackage{amssymb}
\usepackage{graphicx}
\usepackage[dvipdfm]{hyperref}

\newcommand{\ymax}{\bar y}
\newcommand{\pauli}{\sigma}

\begin{document}

\title{Probability distribution of Majorana end-state energies in disordered wires}

\author{Piet W.\ Brouwer} 
\affiliation{\mbox{Dahlem Center for Complex Quantum Systems and Fachbereich Physik, Freie Universit\"at Berlin, 14195 Berlin, Germany}}

\author{Mathias Duckheim} 
\affiliation{\mbox{Dahlem Center for Complex Quantum Systems and Fachbereich Physik, Freie Universit\"at Berlin, 14195 Berlin, Germany}}

\author{Alessandro Romito}
\affiliation{\mbox{Dahlem Center for Complex Quantum Systems and Fachbereich Physik, Freie Universit\"at Berlin, 14195 Berlin, Germany}}

\author{Felix von Oppen}
\affiliation{\mbox{Dahlem Center for Complex Quantum Systems and Fachbereich Physik, Freie Universit\"at Berlin, 14195 Berlin, Germany}}

\date{\today}

\begin{abstract}
One-dimensional topological superconductors harbor Majorana bound
states at their ends. For superconducting wires of finite length $L$,
these Majorana states combine into fermionic excitations with an
energy $\varepsilon_0$ that is exponentially small in $L$. Weak
disorder leaves the energy splitting exponentially small, but affects
its typical value and causes large sample-to-sample fluctuations. We
show that the probability distribution of $\varepsilon_0$ is log
normal in the limit of large $L$, whereas the distribution of the
lowest-lying bulk energy level $\varepsilon_1$ has an algebraic tail
at small $\varepsilon_1$. Our findings have implications for the speed
at which a topological quantum computer can be operated.
\vspace*{-.5cm}\\
\end{abstract}

%{74.78.Na,73.63.Nm,03.67.Lx,71.23.-k}
%71.23.-k Electronic structure of disordered solids
%03.67.Lx Quantum computation architectures and implementations 
%73.63.Nm Quantum wires
\maketitle

{\em Introduction.}---Quantum bits based on topologically protected states promise a platform for error-free quantum computation \cite{kn:kitaev2003,kn:freedman1998,kn:nayak2008}. Since information is stored in states with a topologically protected degeneracy, qubits that rely on this principle are immune to local external sources of decoherence. In practical realizations, however, the energy splitting of the topological qubit is not exactly zero, because of finite-size effects. This poses a restriction on the speed at which a quantum computer must be operated: Operations have to be performed in a time that is short in comparison to the inverse energy splitting, but long in comparison to the inverse excitation gap for (non-topological) excitations.

A particularly promising possibility to realize topologically protected zero-energy states is found in one-dimensional spinless $p$-wave superconductors, which are known to have zero-energy Majorana fermion states at their ends \cite{kn:kitaev2001,kn:motrunich2001}. Although Majorana excitations are insufficient to build a universal topological quantum computer, their implementation may considerably reduce the minimum required accuracy of qubit operations. There are several proposals for the experimental realization of such topological superconducting wires \cite{kn:sau2010,kn:lutchyn2010,kn:oreg2010,kn:potter2010,kn:duckheim2011,kn:chung2011}. In some of these, one-dimensional wires can be brought into an alternation of topological and non-topological domains, with Majorana bound states at the domain boundaries \cite{kn:fu2008,kn:lutchyn2010,kn:oreg2010}, while the location of the Majorana states can be controlled via gate voltages or magnetic fields \cite{kn:oreg2010,kn:alicea2011}.

Experimental realizations necessarily involve topological domains of finite length $L$, as well as disorder. For finite $L$, the Majorana end states fuse into fermionic excitations at a finite energy $\varepsilon_{0}$ that is exponentially small in $L/\xi$, where $\xi$ is the superconductor coherence length \cite{kn:motrunich2001}. In disordered wires, this sets a lower bound for the speed of qubit operations which is sample specific. 
On the other hand, disorder is known to cause a Lifschitz tail of localized states below the gap $\Delta$ \cite{kn:motrunich2001,kn:gruzberg2005}. 
Since operations with Majorana states require that they are transported through the quantum wires \cite{kn:alicea2011}, the lowest-lying bulk state of energy $\epsilon_1$ provides an upper bound for the speed of qubit operations. In view of possible experimental applications and their limitations, it is essential to know the full probability distribution of the energies $\varepsilon_0$ and $\varepsilon_1$. This problem is addressed in this paper.

{\em Solitons in Dirac equation with random mass.---}We first consider the Dirac Hamiltonian with random mass, 
\begin{equation}
  H = v_{\rm F} p \pauli_z - \Delta(x) \pauli_x - V(x) \pauli_x,
  \label{eq:model2}
\end{equation}
as a simple model for a topological superconductor with Majorana end
states. Here $v_{\rm F}$ is the Fermi velocity, $\Delta$ the effective
superconducting gap, and $\pauli_x$ and $\pauli_z$ are Pauli
matrices. The disorder potential $V(x)$ is taken according to a
Gaussian distribution with zero mean and correlator $\langle V(x)
V(x') \rangle = \gamma \delta(x-x')$, corresponding to the mean free
path $l = v_{\rm F} \tau = v_{\rm F}^2/\gamma$ in the normal
state. (Long-range correlated disorder which breaks the system into
topological and nontopological regions has been considered in Ref.\
\onlinecite{kn:shivamoggi2010,kn:flensberg2010}.) The Hamiltonian
(\ref{eq:model2}) arises as a low-energy effective Hamiltonian for
semiconductor wires with strong spin-orbit coupling in proximity to a
conventional superconductor and in the presence of a magnetic field
\cite{kn:brouwer2011}. In that case, the effective gap $\Delta$ in
Eq.\ (\ref{eq:model2}) is the difference between the applied magnetic
field and the proximity-induced superconducting gap in the absence of
a magnetic field. The same Hamiltonian arises in a number of other
contexts, such as fermions on a lattice with random hopping amplitudes
\cite{kn:eggarter1977theodorou1976}, narrow-gap semiconductors
\cite{kn:keldysh1963,kn:ovchinnikov1977}, or organic molecules
\cite{kn:kim1993}. What appears here as a pair of Majorana end states
is referred to as a ``soliton--anti-soliton pair'' in these contexts
\cite{kn:heeger1988}.

We describe a topological domain of length $L$ by setting $\Delta(x) \to -\infty$ for $x < 0$ and $x > L$ and $\Delta(x) = \Delta$ for $0 < x < L$. Since the system is fully gapped for $x < 0$ and $x > L$, any states contributing to the integrated density of states $N(\varepsilon)= \int_0^{\varepsilon} d\varepsilon' \nu(\varepsilon')$ must be localized in the topological domain at $0 < x < L$. In the presence of disorder, $N(\varepsilon)$ is a random quantity with probability distribution $P_{\varepsilon}(N)$, which is related to the probability distribution $p_j(\varepsilon)$ of the energy level $\varepsilon_j$ ($j=0,1,2,\ldots$) through the equalities
\begin{equation}
  p_j(\varepsilon) = - \frac{\partial}{\partial \varepsilon} \sum_{j'=0}^{j} P_{\varepsilon}(j').
  \label{eq:pj}
\end{equation}
Thus the distribution functions of the Majorana end-state energy $\epsilon_0$ and of the lowest bulk state energy $\varepsilon_1$ obey $p_0(\varepsilon) = - {\partial P_{\varepsilon}(0)}/{\partial \varepsilon}$ and $p_1(\varepsilon) = - {\partial [P_{\varepsilon}(0) + P_{\varepsilon}(1)]}/{\partial \varepsilon}$, respectively. 

For each disorder configuration, we can calculate the energy levels $\varepsilon_j$ from the scattering matrix $S(\varepsilon,x')$ of a wire with Hamiltonian (\ref{eq:model2}) for $x < x'$ and $H = v_{\rm F} p \pauli_z$ for $x > x'$. The relation between $S$ (which is really a complex number of unit modulus in the present case) and the energy levels is given by the Friedel sum rule, 
\begin{equation}
  N(\varepsilon) = (2 \pi i)^{-1} \lim_{x' \to \infty} \ln \det [S(\varepsilon,x') S(\varepsilon,-x')^{-1}],
  \label{eq:Friedel}
\end{equation}
where $\lim_{x' \to -\infty} S(\varepsilon,x') = i$. The probability distributions $P_{\varepsilon}(j')$ can then be obtained by considering the evolution of $S(\varepsilon,x^\prime)$ upon repeatedly increasing $x'$ by the small amount $\delta x' \ll \min(l,\xi)$, $\xi = v_{\rm F}/\Delta$ being the superconducting coherence length in the topological domain. 

\begin{figure}
% \epsfxsize=.95\hsize
% \epsffile{fig-1.eps}
\includegraphics[width = 1.0 \linewidth]{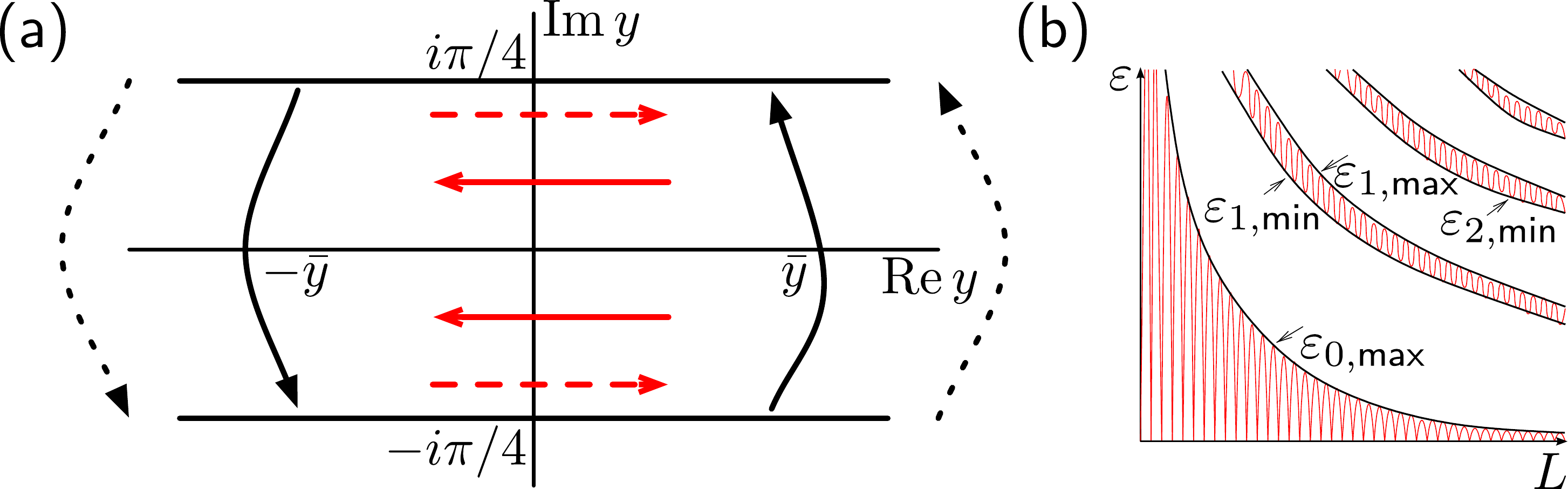}
%\hspace{0.05\hsize}
%\epsfxsize=0.35\hsize
%\epsffile{fig3.eps}
\caption{\label{fig:1} (color online) (a) The variable $y$ in the
  parametrization (\ref{eq:Sy}) of the scattering matrix takes values
  on the real axis $\pm i \pi/4$. The black dotted arrows indicate the
  boundary conditions at $y=\pm \infty$.  In the simplified Langevin
  process used for the asymptotic analysis (\ref{eq:ydiffusion}), the
  boundary conditions are at $\mbox{Re}\, y = \pm \ymax$ (full black
  arrows).  The direction of the drift term in topological ($0<x'<L$)
  and non-topological ($x'<0$ and $x'>L$ ) regions is indicated by the
  full red arrow and dashed red arrow, respectively.  (b) Typical
  dependence of the Majorana end-state energy $\varepsilon_0$ and the
  energies $\varepsilon_1$, $\varepsilon_2$, \ldots, of low-lying bulk
  states on the system size $L$% for a given disorder realization
  .}
%\caption{\label{fig:1} FIXFIXFIXFIX(a) The variable $y$ in the parametrization (\ref{eq:Sy}) of the scattering matrix takes values on the real axis $\pm i \pi/4$. The arrows indicate the boundary conditions at $y=\pm \infty$. (b) In the simplified Langevin process used for the asymptotic analysis, the variables are restricted to the range $|\mbox{Re}\, y| < \ymax$. (c) Typical dependence of the Majorana end state energy $\varepsilon_0$ and the energies $\varepsilon_1$, $\varepsilon_2$, \ldots, of low-lying bulk states on the system size $L$ for a given disorder realization.}
\end{figure}

The evolution of $S(\varepsilon,x')$ takes the form of a Langevin process \cite{kn:beenakker1997}. This Langevin process takes its simplest form if we use the parametrization
\begin{equation}
  S = i \tanh y, \label{eq:Sy}
\end{equation}
instead of the standard parametrization of $S$ through the scattering phase, as the parametrization (\ref{eq:Sy}) makes the noise term become independent of $S$. The parameter $y$ takes values on the real axis $\pm i \pi/4$, see Fig.\ \ref{fig:1}a, and is continuous at $y = \pm \infty$. Concatenating the scattering matrix for a given $x^\prime$ with the scattering matrix of an added slice of length $\delta x^\prime$, we can compute the corresponding change $\delta y$ of $y$. This yields the Langevin process 
 \begin{equation}
  \langle \delta y \rangle = 
  \frac{\delta x'}{v_{\rm F}}
  \left[ i \varepsilon  \sinh 2 y - \Delta(x') \right],\ \
  \langle \delta y^2 \rangle = \frac{\delta x'}{l}.
  \label{eq:langevin2}
\end{equation}
Near $y = \pm \infty$, the shifts are dominated by the term proportional to $\varepsilon$, which unidirectionally couples the branch at $y = \pm \infty \mp i \pi/4$ into the branch at $y = \pm \infty \pm i \pi/4$, see Fig.\ \ref{fig:1}. The initial condition for $x' < 0$ is $y = \infty + i \pi/4$. 
For $x' > L$ the Langevin process returns $y$ to the starting point $y = \infty + i \pi/4$. The return takes place via the upper branch at $\mbox{Im}\, y = \pi/4$ if $\mbox{Im} y(L) = \pi/4$, and via the lower branch otherwise.

The Friedel sum rule (\ref{eq:Friedel}) now identifies
$P_{\varepsilon}(N)$ as the probability distribution of the number of
times $N$ that the variable $y$ has passed through the point at
$-\infty$ upon increasing $x'$ from $x'=0$ to $x'=L$. We have
calculated this probability distribution through direct numerical
simulation of the Langevin process, as well as through an asymptotic
analysis valid in the limit $\varepsilon \ll \min(\Delta,1/\tau)$.

For the asymptotic analysis, we observe that for $\varepsilon \ll \Delta$ the Langevin process (\ref{eq:langevin2}) is dominated by the term proportional to $\varepsilon$ if $|\mbox{Re}\, y| \gtrsim \ymax(\varepsilon)$, with 
\begin{equation}
  \ymax(\varepsilon) = \frac{1}{2}
  \ln \frac{\max(2 \Delta,1/\tau)}{\varepsilon},
  \label{eq:ymax}
\end{equation}
whereas the other terms dominate if $|\mbox{Re}\, y| \lesssim \ymax(\varepsilon)$. To logarithmic accuracy, we approximate the Langevin process (\ref{eq:phi}) by a truncated process in which $|\mbox{Re}\, y| < \ymax$ and the energy contribution to $\langle \delta y \rangle$ is omitted \cite{kn:brouwer2000b}. The unidirectional connection between the upper and lower branches now takes place at $\mbox{Re}\, y = \pm \ymax(\varepsilon)$, see Fig.\ \ref{fig:1}b. The resulting Langevin process for $y$ is then specified by the equations
\begin{equation}
  \langle \delta y \rangle = - \delta x'/\xi,\ \
  \langle \delta y^2 \rangle = \delta x'/2 l,
  \label{eq:ydiffusion}
\end{equation}
with absorbing boundary conditions (``sink'') at $y=\pm(\ymax(\varepsilon) - i \pi/4)$ and hard wall boundary conditions at $y=\pm(\ymax(\varepsilon) + i \pi/4)$, see Fig.\ \ref{fig:1}b. 

\begin{figure}
% \epsfxsize=0.99\hsize
% \epsffile{fig-2.eps}
\includegraphics[width = 1.0 \linewidth]{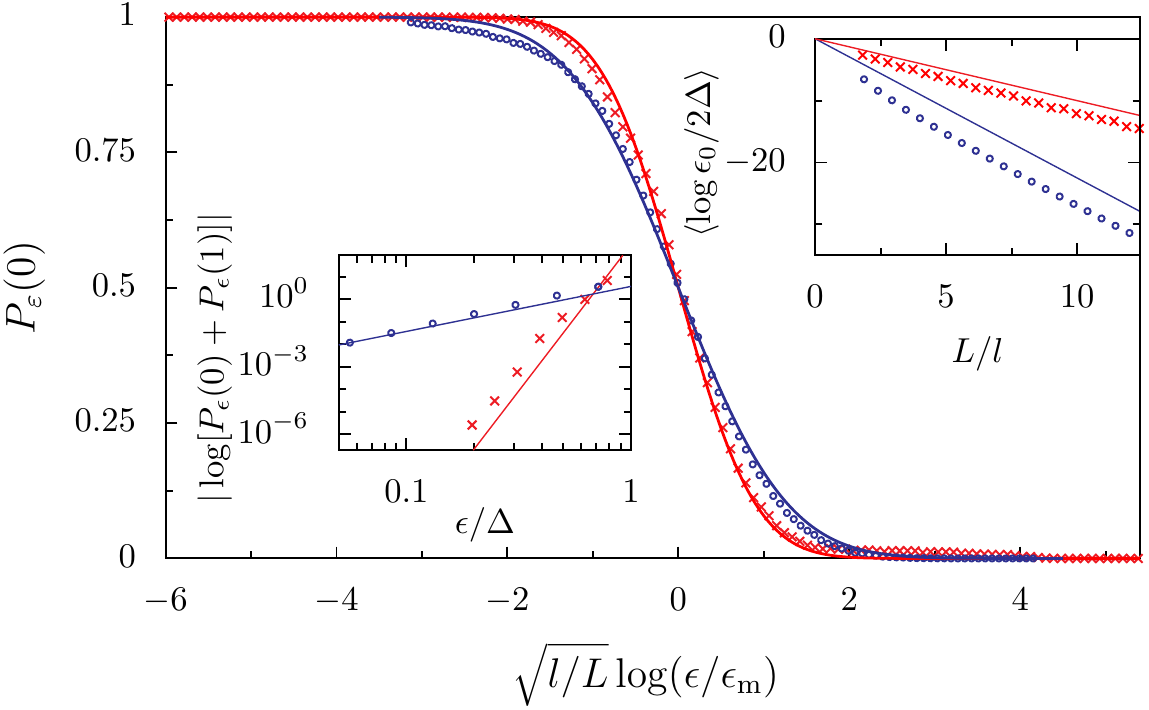}
\caption{\label{fig:2} (color online) Integrated probability density $P_\epsilon(0)$
  of the Majorana end-state energies $\varepsilon_{0}$ (blue dots) and
  $\varepsilon_{0,\rm max}$ (red crosses) in the two models obtained from
  a numerical solution of the Langevin process, % in
  % Eq.~(\ref{eq:langevin2}) (blue dots) and Eq.~(\ref{eq:phi}) (red
  % dots)
  together with the theoretical prediction (solid). Energies
  are normalized to the median $\epsilon_{\rm m}$ of the
  distribution.  Left inset: Logarithm of the integrated probability
  distribution $P_{\varepsilon}(0)+P_{\varepsilon}(1)$ of the lowest
  lying bulk state energies $\varepsilon_{1}$ and $\varepsilon_{1,\rm
    min}$. Right inset: Average of $\log \varepsilon_{0}$ and $\log
  \varepsilon_{0, {\rm max}}$ vs. length. In both insets the slope of the solid lines is
  given by the theoretical predictions of the main text.  }
\end{figure}

There is a qualitative difference between the Langevin processes at
the upper branch ($\mbox{Im}\, y = \pi/4$) and the lower branch
($\mbox{Im}\, y = -\pi/4$): At the upper branch, the drift term
proportional to $1/\xi$ pushes the variable $y$ towards the sink,
whereas at the lower branch it keeps it away from the sink. The slow
diffusion in the latter case does not affect $\varepsilon_0$, but it
dominates the probability distribution of all higher levels
$\varepsilon_j$, $j=1,2,\ldots$. By analyzing the diffusion process on
the upper branch, we find that the probability $P_{\varepsilon}(0)$ is
\cite{kn:smoluchowski1916}
\begin{equation}
  P_{\varepsilon}(0) = 
  \frac{1}{2}\, \mbox{erfc} \left(
  \frac{ L/\xi - 2 \ymax(\varepsilon) 
  }{\sqrt{2L/l}} \right).
  \label{eq:P0maj}
\end{equation}
Recalling that $\ymax(\varepsilon)$ is given by Eq.\ (\ref{eq:ymax}), we conclude that $\ln (\varepsilon_{0}/2\Delta)$ has a normal distribution with mean and variance given by
\begin{eqnarray}
  \langle \ln (\varepsilon_{0}/2\Delta) \rangle &=& - L/\xi,\nonumber \\
  \mbox{var}\,  \ln (\varepsilon_{0}/2\Delta) &=& L/l,
  \label{eq:lognormal2}
\end{eqnarray}
up to corrections of order unity that can not be determined from the
above argument. Similarly, analyzing the diffusion process on the
lower branch, we find $P_{\varepsilon}(0) + P_{\varepsilon}(1) =
e^{-\langle N(\varepsilon) \rangle}$, where the disorder-averaged
integrated density of states $\langle N(\varepsilon) \rangle \propto
(L/\xi) (\varepsilon/\Delta)^{2 l/\xi}$
\cite{kn:ovchinnikov1977,kn:brouwer2000b}, with proportionality
constant that could not be determined from the asymptotic
analysis. From this, we find that $p_1(\varepsilon) \propto (L
\tau/\xi) (\varepsilon/\Delta)^{2 l/\xi - 1}$ for small
$\varepsilon$. The theoretical predictions are compared to numerical
simulations of the Langevin process in Fig.\ \ref{fig:2}.

{\em One-dimensional spinless $p$-wave superconductor.}---We now extend these results to a one-dimensional $p$-wave superconductor, using a continuum version of the model considered in Refs.\ \onlinecite{kn:motrunich2001,kn:kitaev2001},
\begin{equation}
  H = \left( \frac{p^2}{2 m} + V(x) - \mu \right) \pauli_z
  - \Delta' p \pauli_x.
  \label{eq:pwave}
\end{equation}
Here $\mu$ is the chemical potential, $V(x)$ the disorder potential, which we take according to the same distribution as in the previous case, and $\pauli_z$ and $\pauli_x$ are Pauli matrices acting in the electron-hole space. We model a topological domain of finite length $L$ by setting $\mu = -\infty$ for $x < 0$ and $x > L$, and $\mu = p_{\rm F}^2/2m > 0$ for $0 < x < L$, $p_{\rm F} = m v_{\rm F}$ being the Fermi momentum \cite{kn:read2000}. Further, $\Delta = \Delta' p_{\rm F}$ is the superconducting gap. Throughout our calculation we assume that 
$\xi$, $l \gg \hbar/p_{\rm F}$.

Previous studies of lattice versions of the model (\ref{eq:pwave}) addressed the disorder-averaged density of states $\langle \nu(\varepsilon) \rangle$ in the limit $L \to \infty$ \cite{kn:motrunich2001,kn:gruzberg2005}. Using a strong-disorder renormalization group approach, Motrunich {\em et al.} showed that the model (\ref{eq:pwave}) is in a topological phase if the disorder strength is below a critical value, and in a non-topological phase for stronger disorder. On both sides of the critical disorder strength, the density of states $\nu(\varepsilon)$ has a power law dependence on $\varepsilon$ for $\varepsilon \ll \Delta$, with an exponent that depends on the disorder strength. For the continuum model (\ref{eq:pwave}) we now show that the transition is at $\xi = 2l$ and calculate the probability densities of the Majorana end-state energy $\varepsilon_0$ and the lowest-lying bulk level for disorder strengths below the critical value.

Our calculation essentially follows the approach taken above for the Dirac equation with random mass with some modifications. We define a $2\times 2$
scattering matrix $S(\varepsilon,x')$ of a wire with Hamiltonian given by Eq.\ (\ref{eq:pwave}) for $x < x'$ and by $H = (p^2/2m) \pauli_z$ for $x > x'$ and parametrize $S$ through
\begin{equation}
  S(\varepsilon,x') = -\frac{1}{2} \sum_{\pm} \left( \begin{array}{cc}
  \pm e^{i \phi} \tanh y_{\pm} &
  -i \tanh y_{\pm} \\
  i \tanh y_{\pm} &
  \pm e^{-i \phi} \tanh y_{\pm} \end{array} \right),
  \label{eq:Sparam}
\end{equation}
where the variables $y_{+}$ and $y_{-}$ take values on the real axis $\pm i \pi/4$, see Fig.\ \ref{fig:1}, and $\phi$ is a real phase. The energy levels can no longer be calculated from the Friedel sum rule (\ref{eq:Friedel}), but instead have to be obtained from the condition $\det[1 + S(\varepsilon,L)] = 0$, which becomes 
\begin{equation}
  \cos \phi = \coth(y_--y_+).
  \label{eq:statecond}
\end{equation}
in the parametrization (\ref{eq:Sparam}). For $\hbar/p_{\rm F} \ll \delta x' \ll l,\xi$, the resulting Langevin processes for the variables $y_{\pm}$ and the phase $\phi$ decouple, 
\begin{eqnarray}
  \langle \delta y_{\pm} \rangle &=&
  \frac{\delta x'}{v_{\rm F}}
  \left(i \varepsilon \sinh 2 y_{\pm} - \Delta \right)
  + \frac{\delta x'}{2 l} \coth(y_{\pm} - y_{\mp}), \nonumber \\
  \langle \delta y_{\pm}^2 \rangle &=& -\langle \delta y_{\pm} \delta y_{\mp} \rangle \,= \,{\delta x'}/{2 l} ,
   \nonumber\\
  \langle \delta \phi \rangle &=& 2 p_{\rm F} \delta x', \nonumber \\
  \langle \delta \phi^2 \rangle &=&
  {4 \delta x'}/{l} + ({\delta L}/{2 l}) \coth^2(y_+-y_-).
  \label{eq:phi}
\end{eqnarray}
The initial condition is
\begin{equation}
  \phi(0) = 0\ \ \mbox{and}\ \ y_{\pm}(0) = \pm \infty \pm i\pi/4.
  \label{eq:initial}
\end{equation}

For a given disorder realization, the solutions $\varepsilon_j$ of
Eq.\ (\ref{eq:statecond}) oscillate as a function of $L$, with
oscillation period $\approx \pi/p_{\rm F}$, as shown schematically in
Fig.\ \ref{fig:1}b. This follows from the observation that $y_+$ and
$y_-$ are ``slow'' as a function of $L$ [they vary on the scale
$\min(l,\xi)$], whereas $\phi$ is a ``fast'' variable ($\delta
\phi/\delta x' \approx 2 p_{\rm F}$): Starting from the initial
condition (\ref{eq:initial}), solutions of Eq.\ (\ref{eq:statecond})
then appear in quick succession upon increasing $L$ at fixed
$\varepsilon$, until $y_+$ passes through the point at $-\infty$ or
$y_-$ passes through the point at $+\infty$, whichever occurs
first. No solutions of Eq. (\ref{eq:statecond}) are found upon
increasing $L$ further, until eventually again one of the variables
$y_{\pm}$ passes through a point at infinity and solutions to Eq.\
(\ref{eq:statecond}) reappear in quick succession, cp.\ Fig.\
\ref{fig:1}b.

We now calculate the probability distributions $p_{0,{\rm max}}$ and $p_{1,{\rm min}}$ of the {\em maximum} $\varepsilon_{0,{\rm max}}$ and the {\em minimum} $\varepsilon_{1,{\rm min}}$ with respect to variations of $L$ of order $\pi/p_{\rm F}$. (Note that $\varepsilon_{0,{\rm max}}$ and $\varepsilon_{1,{\rm min}}$ are the energies relevant for setting the operation speed of a hypothetical topological quantum computer.)
Repeating the arguments of the first part of this paper, these probabilities obey
\begin{eqnarray}
  p_{j,{\rm max}}(\varepsilon) &=& -\frac{\partial}{\partial \varepsilon} \sum_{j'\le 2j} 
  P_{\varepsilon}(j'),\\
  p_{j,{\rm min}}(\varepsilon) &=& -\frac{\partial}{\partial \varepsilon} \sum_{j'\le 2j-1} 
  P_{\varepsilon}(j'),
\end{eqnarray}
where $P_{\varepsilon}(N)$ is the probability that (in total) the
variables $y_{+}$ or $y_{-}$ have passed $N$ times through the points
at $\pm \infty$ upon increasing $x'$ from $0$ to $L$. We have
calculated these probabilities from direct numerical simulation of the
Langevin process, as well from an asymptotic analytical solution valid
in the limit $\varepsilon \ll \min(\Delta,1/\tau)$. For the asymptotic
analysis, we make the same simplification of the Langevin process as
in the case of the Dirac equation with random mass. In addition, we
observe that for the energies of interest, one of the variables
$y_{\pm}$ effectively remains pinned at $-\ymax(\varepsilon) - i
\pi/4$, so that the factor $\coth(y_+-y_-)$ in the interaction term
may be approximated by $\pm 1$. The resulting Langevin process for the
remaining variable is then specified by the equations
\begin{equation}
  \langle \delta y \rangle = - \delta L/\xi + \delta L/2 l,\ \
  \langle \delta y^2 \rangle = \delta L/2 l,
  \label{eq:yplusdiffusion}
\end{equation}
with the boundary conditions as specified below Eq.\ (\ref{eq:ydiffusion}). The result for $P_{\varepsilon}(0)$ has the same functional form as in the case of the random-mass Dirac equation, and we conclude that $\ln (\varepsilon_{0,{\rm max}}/2\Delta)$ has a normal distribution with mean and variance given by
\begin{eqnarray}
  \langle \ln (\varepsilon_{0,{\rm max}}/2\Delta) \rangle &=& - L[1/\xi - 1/(2 l)],\nonumber \\
  \mbox{var}\,  \ln (\varepsilon_{0,{\rm max}}/2\Delta) &=& L/2l,
  \label{eq:lognormal}
\end{eqnarray}
up to corrections of order unity that can not be determined from the asymptotic argument. The end-state energy remains exponentially small in $L$ as long as $2l > \xi$, which identifies $2 l = \xi$ as the critical disorder strength that drives the system into the non-topological phase. Similarly, we find $P_{\varepsilon}(0) + P_{\varepsilon}(1) = e^{-\langle N(\varepsilon) \rangle}$, with $\langle N(\varepsilon) \rangle \propto (L/\xi) (\varepsilon/\Delta)^{4 l/\xi - 2}$ \cite{kn:brouwer2000b,kn:motrunich2001,kn:gruzberg2005}, from which we conclude that $p_{1,{\rm min}}(\varepsilon) \propto (L \tau/\xi) (\varepsilon/\Delta)^{4l/\xi - 3}$ for small energies. At the critical disorder strength, the integrated density of states takes the Dyson form $N(\varepsilon,L) \propto \ln^2(\varepsilon/\Delta)$ \cite{kn:motrunich2001,kn:titov2001,kn:gruzberg2005}. The theoretical prediction for $p_{0,{\rm max}}$ is compared to numerical simulations of the Langevin process in Fig.\ \ref{fig:2}.

{\em Conclusions.---}For both models of a topological superconducting wire, we find that the energy splitting $\epsilon_0$ of the Majorana end states has a log-normal distribution, implying large sample-to-sample fluctuations. Nevertheless, for sufficiently long wires, the width of the log-normal distribution remains small compared to its average. In this case, the lower limit on the speed of the qubit operations is well determined by the typical value of the log-normal distributions in Eqs.\ (\ref{eq:lognormal2}) and (\ref{eq:lognormal}), which is exponentially small in $L/\xi$. By contrast, we find that the energy $\varepsilon_1$ of the lowest-lying bulk state is algebraically small in $L/\xi$. This implies that in principle, there is a large parameter window in which both conditions on the operation speed can be met if $L$ is made sufficiently large. It is important to note, however, that with increasing disorder or increasing $L$, this parameter window is shifted to lower energies which would require the topological quantum computer to operate at a lower temperature and lower speed.

This work is supported by the DFG through SPP 1285 and the Alexander von Humboldt Foundation.

\end{document}